# Identifying conserved protein complexes between species by constructing interolog networks

**Phi Vu Nguyen [1], Sriganesh Srihari [2] and Hon Wai Leong [1§]**

[1]Department of Computer Science, National University of Singapore, Singapore 117590

[2]Institute for Molecular Bioscience, The University of Queensland, St. Lucia, QLD 4072, Australia

§Corresponding author

Email addresses:

    PVN: nphivu@comp.nus.edu.sg

    SS: sriganesh.m.s@gmail.com

    HWL: leonghw@comp.nus.edu.sg

# Abstract

**Background**

Protein complexes *conserved* across species indicate processes that are *core* to cellular machinery (*e.g.* cell-cycle or DNA damage-repair complexes conserved across human and yeast). While numerous computational methods have been devised to identify complexes from the protein interaction (PPI) networks of individual species, these are severely limited by noise and errors (false positives) in currently available datasets. Our analysis using human and yeast PPI networks revealed that these methods missed several important complexes including those conserved between the two species (*e.g.* the MLH1-MSH2-PMS2-PCNA mismatch-repair complex). Here, we note that much of the functionalities of yeast complexes have been conserved in human complexes not only through sequence conservation of proteins but also of critical *functional domains*. Therefore, integrating information of domain conservation might throw further light on conservation patterns between yeast and human complexes.

**Results**

We identify conserved complexes by constructing an *interolog network* (IN) leveraging on the *functional conservation* of proteins between species through *domain conservation* (from Ensembl) in addition to sequence similarity. We employ 'state-of-the-art' methods to cluster the interolog network, and map these clusters back to the original PPI networks to identify complexes conserved between the species. Evaluation of our IN-based approach (called COCIN) on human and yeast interaction data identifies several additional complexes (76% recall) compared to direct complex detection from the original PINs (54% recall). Our analysis revealed that the IN-construction removes several non-conserved interactions many of which are false positives, thereby improving complex prediction. In fact removing non-conserved interactions from the original PINs also resulted in higher number of conserved complexes, thereby validating our IN-based approach. These complexes included the mismatch repair complex, MLH1-MSH2-PMS2-PCNA, and other important ones namely, RNA polymerase-II, EIF3 and MCM complexes, all of which constitute core cellular processes known to be conserved across the two species.

**Conclusions**

Our method based on integrating domain conservation and sequence similarity to construct interolog networks helps to identify considerably more conserved complexes between the PPI networks from two species compared to direct complex prediction from the PPI networks. *Availability*: http://www.comp.nus.edu.sg/~leonghw/COCIN/

# Background

Complexes of physically interacting proteins form fundamental units responsible for driving key biological processes within cells. Even in the simple model organism *Saccharomyces cerevisae* (budding yeast), these complexes are composed to several protein subunits that work in a coherent fashion to carry out cellular functions. Therefore a faithful reconstruction of the entire set of complexes (the 'complexosome') from the set of physical interactions (the 'interactome') is essential to understand their organisation and functions as well as their roles in diseases [1-4].

In spite of the significant progress in computational identification of protein complexes from protein interaction (PPI) networks over the last few years (see the surveys [1,2]), computational methods are severely limited by noise (false positives) and lack of sufficient interactions (*e.g.* membrane-protein interactions) in currently available PPI datasets, particularly from human, to be able to completely reconstruct the complexosome [1,2]. For example, several complexes involved in core cellular processes such as cell cycle and DNA damage response (DDR) are not present in a recent (2012) compendium of human protein complexes (http://human.med.utoronto.ca/) assembled solely by computational identification of complexes from high-throughput PPIs [5]; a web-search (as of Feb 2013) in this compendium for BRCA1 does not yield any complexes even though BRCA1 is known to participate in three fundamental complexes in DDR *viz.* BRCA1-A, BRCA1-B and BRCA1-C complexes [6-8]. A possible reason for missing these complexes is the lack of sufficient PPI data required for identifying them even using the best available algorithms. But, the authors of this compendium note that many human complexes appear to be ancient and slowly evolving – roughly a quarter of the predicted complexes overlapped with complexes from yeast and fly, with half of their subunits having clear orthologs [5]. Therefore, it is useful to devise effective computational methods that look for evidence from evolutionary conservation to complement PPI data to reconstruct the full set of complexes.

In the attempt to integrate evolutionary information with PPI networks, Kelley *et al.* [9] and Sharan *et al.* [10] devised methods to construct an *orthology graph* of conserved interactions from two species, which in their experiments were yeast (*S. cerevisae*) and bacteria (*H. pylori*), using a sequence homology-based (using BLAST E-score similarity) mapping of proteins between the species. Dense sub-graphs induced in this orthology graph represented putative complexes conserved between the two species. The complexes so-identified were involved in core cellular processes conserved between the two species – e.g. those in protein translation, DDR and nuclear transport. Van Dam and Snel (2008) [11] studied rewiring of protein complexes between yeast and human using high-throughput PPI datasets mapped onto known yeast and human complexes. From their experiments, they concluded that a majority of co-complexed protein pairs retained their interactions from yeast to human indicating that the evolutionary dynamics of complexes was not due to extensive PPI network rewiring within complexes but instead due to gain or loss of protein subunits from yeast to human. Hirsh and Sharan [12] developed a protein evolution-based model and employed it to identify conserved protein complexes between yeast and fly, while Zhenping *et al.* [13] used integer quadratic programming to align and identify conserved regions in molecular networks. Marsh *et al.* [14] integrated data on PPI and structure to understand mechanisms of protein conservation; they found that during evolution gene fusion events tend to optimize complex assembly by simplifying complex topologies, indicating genome-wide pathways of complex assembly.

**Integrating domain conservation**

Inspired from these works, here we devise a novel computational method to identify *conserved complexes* and apply it to yeast and human datasets. A crucial point we note on the conservation from yeast to human is that many cellular mechanisms, though conserved, have in fact evolved many-fold in complexity – for example, cell cycle and DDR. Consequently, while several proteins in these mechanisms are conserved by sequence similarity (e.g. RAD9 and hRAD9), there are others that are unique (non-conserved) to human (e.g. BRCA1); see **Figure 1.** These non-conserved proteins perform similar functions (e.g. cell cycle and DDR) as their conserved counterparts, but do not show high sequence similarity to any of the yeast proteins. A deeper examination reveals that these proteins in fact contain *conserved functional domains* – for example, the BRCT domain which is present in yeast RAD9 and human hRAD9 is also present in the non-conserved human BRCA1 and 53BP1; all of these play crucial roles in DDR [15]. Similar structure can be seen in the case of RecQ helicases – several helicase domains are conserved from the yeast SGS1 to human BLM and WRN, but there are three helicases RECQ1,4,5 which are unique to human that also contain these helicase domains [16]. Therefore, integrating information on *functional conservation*, mainly through *domain conservation*, can help to identify considerably more (functionally) conserved complexes than mere sequence similarity, thereby throwing further light on the conservation patterns of complexes in particular and cellular processes in general.

In order to achieve this, simple BLAST-based scores as used in earlier works [9-13] to measure homology between yeast and human proteins do not suffice. Here, we integrate multiple databases including Ensembl [17] and OrthoMCL [18] to build homology relationships among proteins; these databases use a variety of information to construct *orthologous groups* among proteins including checking for *conserved domains*. The integration of these databases generates *many-to-many* correspondence between yeast and human proteins instead of the predominantly one-to-one correspondence obtained by from BLAST-based similarity.

We devise a novel computational method to construct an *interolog network* using domain information along with PPI conservation between human and yeast. Next, we identify dense clusters within the interolog network using current 'state-of-the-art' PPI-clustering methods (as against traditional clustering methods used in [9,10]). These clusters when mapped back to the PPI networks reveal conserved dense regions, many of which correspond on conserved complexes.

Our experiments here reveal that,

(i) integrating domain information generates many valuable interactions from the many-to-many ortholog relationships in the interolog network, thereby enhancing its quality;

(ii) interolog network also reduces false-positive interactions by accounting for conserved PPIs;

(iii) our interolog network construction aids clustering algorithms to identify far more conserved complexes than direct clustering of the individual PPI networks; and

(iv) many of these conserved complexes are involved in core cellular processes such as cell cycle and DDR throwing further light to the conservation of these cellular processes.

We call our method **COCIN** (COnserved Complexes from Interolog Networks).

## Methods

**Constructing the interolog network**

Given two PPI networks from two species $S_1$ and $S_2$, and the homology information between proteins of the two networks, we construct an *interolog network* $G_I$ as follows. The two PPI networks are represented as $G_1(V_1, E_1)$ and $G_2(V_2, E_2)$, and the homology relationship between the proteins is governed by a *many-to-many correspondence* $\theta: V_1 \rightarrow V_2$. The interolog network is defined as $G_I(V_I, E_I)$, where $V_I = \{v_I = \{p, q\} \mid p \in V_1, q \in V_2, \text{ and } (p, q) \in \theta\}$, and $E_I = \{(v_I, v'_I) \mid v_I = \{p,q\}; v'_I = \{r,s\}; (p, r) \in E_1 \text{ and } (q,s) \in E_2\}$.

Each node in the interolog network represents a *pair of homologous proteins*, one from each species. Each edge in the interolog network represents an interaction that is *conserved* in both species (interolog). However, if a protein $p \in V_1$ can be orthologous to multiple proteins $x \in V_2$ and $x \in V_2$, then we add two vertices to $G_I$ namely $\{p, x\}$ and $\{p, y\}$, and add an edge between two vertices. Doing so integrates the many-to-many relationships obtained due to domain conservation into the interolog network. **Figure 2** below gives a simple example of this network-construction.

Any connected sub-network in this interolog network can be mapped back to conserved sub-networks in the two PPI networks, and this is similar to the orthology graph method introduced by Kelley *et al.* [9] and Sharan *et al.* [10]. However, one unique advantage of our interolog network offers is that we can infer a *collection* of homologous complexes between the species. This property is highly relevant for identifying conserved complexes between yeast and human (revisit **Figure 1**).

In order to achieve this, we integrate multiple databases including Ensembl [17] and OrthoMCL [18] to build our homology relationships among proteins; these databases use a variety of information to construct orthologous groups among proteins including checking for conserved domains.

**Clustering the interolog network and detection of conserved complexes**

We identify dense clusters in the interolog network to detect conserved complexes between the two species. To do this, we tested a variety 'state-of-the-art' PPI network-clustering methods, and found the following three to perform the best – CMC (Clustering by merging Maximal Cliques) by Liu *et al.* [19], MCL (Markov Clustering) by van Dongen [20] and HACO (Hierarchical Clustering with Overlaps) by Wang *et al.* [21]. The comparative assessment of these methods has been confirmed with earlier works [1, 2, 22-24].

CMC operates by first enumerating all maximal cliques in network, and ranks them in descending order of the weighted interaction density. It then iteratively merges highly overlapping cliques to identify dense clusters in the network. MCL simulates a series of random paths (called a flow) and iteratively decomposes the network into a number of dense clusters. HACO performs hierarchical clustering by repeatedly identifying smaller dense clusters and merging these into larger clusters. HACO has an advantage over the traditional hierarchical clustering because it allows for overlaps (protein-sharing) among the clusters.

Upon finding dense clusters in the interolog network, we map back these clusters to sub-networks within the two PPI networks to identify conserved complexes.

**Building a benchmark dataset for conserved protein complexes**

Due to lack of benchmark datasets of conserved protein complexes between human and yeast in the literature, we built our own "gold standard" conserved dataset as follows. Using currently available datasets of manually curated protein complexes of human and yeast, we selected pairs of complexes that shared significant fraction of (homologous) proteins.

For measuring the conservation level of a given complex pair $\{C_1, C_2\}$, where $C_1$ belongs to species $S_1$ and $C_2$ belongs to species $S_2$, we adopted the following *Multi-set Jaccard score*:

*Multi-set Jaccard score*: Let $G_{C1}$ and $G_{C2}$ be the collections of ortholog groups in complexes $C_1$ and $C_2$, respectively. For any group $g_i \in Gc_i$ ($i = 1, 2$), let $I_{Ci}$ represent the multiplicity of the group $g_i$ in complex $C_{i,}$ which essentially is the number of paralogs within the group. Multi-set Jaccard score then given by:

$$MSJ(C_1, C_2) = \frac{\sum_{g_i \in (G_{C1} \cup G_{C2})} \min(I_{C_1}(g_i), I_{C_2}(g_i))}{\sum_{g_i \in (G_{C1} \cup G_{C2})} \max(I_{C_1}(g_i), I_{C_2}(g_i))},$$

There are often duplication of genes (paralogs) within complexes and clusters. Therefore, MSJ takes into account the multiplicity of the groups and does a more conservative and accurate estimation of the conservation between $C_1$ and $C_2$. See **Figure 3** for an illustration.

We selected pairs of complexes that show MSJ ≥ 50% (see result section for details).

## Results

**Preparation of experimental data**

We combined multiple PPI datasets to enhance the coverage of our interactome. We collected PPIs from IntAct [25] (version November 13, 2012) and Biogrid [26] (versions 3.2.95 and 3.2.89) databases for yeast; and from Biogrid [26] and HPRD [27] (Release 9, 2010) for human. **Table 1** and **2** summarise these datasets.

Yeast curated complexes were gathered from Wodak database (CYC2008) [28] and human curated complexes from CORUM (version 09/2009) [29]; these form our benchmark complex datasets (details in **Table 3**). We used Ensembl [17] and OrthoMCL [18] for the homology mapping between human and yeast proteins.

**Criteria for evaluating predicted complexes**

For a predicted complex $C_i$ of one species and a manually curated (benchmark) complex $B_j$, we used Jaccard score based on collections of complex proteins: $J(C_i, B_j) = \frac{|C_i \cap B_j|}{|C_i \cup B_j|}$, which considers $C_i$ a correct prediction for $B_j$ if $J(C_i, B_j) \geq t$, a *match threshold*. We chose $t = 0.50$ in our experiments as suggested by earlier works [19, 22]. $C_i$ is then referred to as a *matched prediction* or *matched predicted complex*, and $B_j$ is referred to as a *derived benchmark complex*.

Based on this, *precision* is computed as the fraction of predicted complexes matching benchmark complexes, and the *recall* is computed as the fraction of benchmark protein complexes covered by our predicted complexes. A correctly predicted complex is also checked against our "gold standard" testing dataset to see if it is a conserved complex, in which case the derived complex is a *derived conserved complex*.

**Results of complex detection using interolog network (IN)**

**Table 4** summarizes the interolog network constructed from yeast and human PPIs. We map back each predicted cluster from the IN to the original PPI networks to predict conserved complexes between the two species.

Firstly, we compared the results of complex detection from COCIN with direct clustering of the original PPI networks using CMC, HACO and MCL as shown in **Tables 5** and **6**. Interestingly, we observed that COCIN, which employs CMC, HACO and MCL for clustering the interolog network, yielded a better recall than these methods on the original PPI networks. Further, because IN capitalises on the existence of interactions in both PPI networks (that is, conservation of interactions), the number of noisy dense clusters in COCIN is considerably reduced thereby enhancing its precision.

**Figure 4** compares a predicted complex $C_i$ through COCIN with two predictions $C_y$ and $C_h$ from the original PPI networks; $C_y$ and $C_h$ form a pair of orthologous complexes, but by direct clustering of the original PPI networks and matching them and not using COCIN. We noticed that $C_y$ and $C_h$ contained several noisy proteins and interactions among them which were false positives. These false positives reduced the Jaccard accuracy of these complexes when matched to known benchmark complexes. We also note that when we computed the complex-derivability index called *Component-Edge score* (this index measures how much of chance a complex can be detected given the topology of a PPI network) proposed in [24], $C_i$ had a higher CE-score compared to $C_y$ and $C_h$ in the networks.

**Figure 5** highlights the improvement of COCIN over CMC, that is, the additional protein complexes of human and yeast detected by COCIN. As many noisy interactions are removed in the IN, among the conserved complexes that are detected by both CMC and COCIN, COCIN on an average obtained higher Jaccard scores.

Some important additional conserved complexes found using COCIN were: RNA Polymerase II, EIF3 complex, MSH2-MLH1-PMS2-PCNA DNA-repair initiation complex, MCM complex, MMR complex, Ubiquitin E3 ligase, transcription factor TFIID, DNA replication factor C, 20S proteasomes (descriptions of these complexes are listed in **Tables 7** and **8**).

**The result of complex detection in the conserved subnetworks**

To further understand the advantage of COCIN on leveraging conservation for better detection of complexes, we performed another experiment *alternative* to the interolog network as follows. We predicted complexes from the *subset of protein interactions of the first species that are conserved in the second* (we call this the *conserved subnetwork* in the first species). However, this can only find complexes of one species at a time, so we map these predicted complexes onto the PPI network of the other species to identify the corresponding conserved complexes. We employed CMC to do clustering on the conserved subnetworks.

Complex prediction from conserved subnetworks showed similar result as COCIN – 16 additional conserved complexes in human and 9 additional conserved complexes in yeast are found. This supported the purpose of IN – to leverage conserved interactions for improving complex prediction.

**Figure 6** shows two other examples that explain why additional conserved complexes are found by COCIN but missed by CMC. We see from this picture that the predicted human complex from IN (the leftmost figure) and the corresponding predicted complex from the conserved subnetwork (the center figure) were contained in a *larger* CMC-predicted complex (the rightmost figure) from the original PPI networks. This larger complex included several noisy proteins that reduce the accuracy of the complex, thereby causing the complex to be missed.

**Comparisons with other complex detection methods in PPI networks**

Similar results were obtained using the other two methods HACO and MCL as well, thereby supporting the effectiveness of COCIN in identifying conserved protein complexes. **Tables 5** and **6** present these comparisons in more details, while **Figures 7** and **8** highlight further substantiate these results.

**Integrating domain information significantly enhances interolog construction**

Finally**, Table 9** summarizes the quality of our testing dataset for conserved protein complexes between yeast and human. We compared the number of benchmark conserved complexes found in both human and yeast using mappings from Ensembl and OrthoMCL under multiple conservation score thresholds (**Figure 9**). *Note* that Ensembl contains homology information based on both sequence similarity as well as domain conservation, while OrthoMCL is predominantly based on sequence similarity. We noticed that using Ensembl homology information can yield more conserved complexes at all conservation score thresholds. Further, **Figure 10** shows that there exist 1-to-many and many-to-many relationships of conservation between human and yeast complexes.

Sharan et al. used whole-sequence similarity to construct the interolog network. Here, we used OrthoMCL as a substitute for the whole-sequence similarity due to technical difficulties of running BLAST for a large number of proteins. We compared the performance of using OrthoMCL against using Ensembl, which uses domain conservation along with sequence similarity to determine orthology. **Table 10** and **Figure 11** show that we obtain an overall improvement in terms of the number of mapped protein pairs, interologs, as well as conserved protein complexes in both human and yeast by incorporating domain information (through Ensembl). This substantiates the improved performance of COCIN over traditional sequence-similarity based methods.

## Conclusions

Identifying conserved complexes between species is a fundamental step towards identification of conserved mechanisms from model organisms to higher level organisms. Current methods based on clustering PPI networks do not work well in identifying conserved complexes, and they are severely limited by lack of true interactions and presence of large amounts of false interactions in existing PPI datasets. Here, we presented a method COCIN based on building interolog networks from the PPI networks of species to identify conserved complexes. Our experiments on yeast and human datasets revealed that our method can identify considerably more conserved complexes that plain clustering of the original PPI networks. Further, we demonstrated that integrating domain information generates many-to-many ortholog relationships which significantly enhances interolog quality and throws further light on conservation of mechanisms between yeast and human.

## Availability

Our COCIN software and the datasets used in this work are freely available at:

http://www.comp.nus.edu.sg/~leonghw/COCIN/

or alternately at: https://sites.google.com/site/mclcaw/

## Acknowledgements

We thank the following funding sources: PVN and LHW are supported by a NUS Singapore ARF grant R252-000-461-112; SS is supported by an Australian NHMRC grant 1028742 to Professor Mark A. Ragan (UQ) and Dr Peter Simpson (UQCCR).

# Figures

**Figure 1 - Conservation of complexes between yeast and human**

Many proteins in yeast have either 'split' into multiple proteins or fused into common proteins in human during evolution. This mechanism is a result of selecting optimal protein assemblies [14] thereby resulting in multi-fold expansion of complexity in human. In order to capture these conservation mechanisms it is necessary to integrate domain along with PPI information.

**Figure 2 - Construction of the interolog network – a simplified example**

Our interolog network constructing integrates PPI and domain conservation information to generate a network that is conducive for clustering algorithms to identify considerably more conserved complexes compared to direct clustering of the original PPI networks from species.

**Figure 3 - Conservation scores for building benchmark complex datasets**

We generate a "gold standard" conserved complexes dataset to test our method. We use two scores here – the Jaccard score for orthologous groups and multi-set Jaccard score.

**Figure 4 - An illustration on a predicted complexes from IN**

(a) A predicted complex in the IN.

(b) The corresponding complex in the human PPI network.

(c) The corresponding complex in the yeast PPI network.

**Figure 5 - COCIN compared to CMC**

COCIN over the interolog network identifies significantly more conserved complexes compared to direct clustering of the original PPI networks using CMC [19].

**Figure 6 - Some examples of additional conserved complexes found in IN**

The clusters detected from the original PPI networks include several noisy proteins and noisy interactions (false positives), thereby reducing their Jaccard accuracies.

**Figure 7 - COCIN compared to HACO**

COCIN over the interolog network identifies significantly more conserved complexes compared to direct clustering of the original PPI networks using HACO [20].

**Figure 8 - COCIN compared to MCL**

COCIN over the interolog network identifies significantly more conserved complexes compared to direct clustering of the original PPI networks using MCL [21].

**Figure 9 - Assessment of Ensembl and OrthoMCL based homology for IN construction and conserved-complex detection**

Ensembl [17] contains protein orthologs based on sequence similarity as well as domain information, while OrthoMCL [18] is predominantly based on sequence similarity. As we can see from the table, using domain information (through Ensembl) generates significantly more many-to-many ortholog mappings thereby enhancing our interolog construction.

**Figure 10 – Some examples of the one-to-many and many-to-many relationships of complex conservation between human and yeast**

Ensembl [17] contains protein orthologs based on sequence similarity as well as domain information, while OrthoMCL [18] is predominantly based on sequence similarity. As we can see from the table, using domain information (through Ensembl) generates significantly more many-to-many ortholog mappings thereby enhancing our interolog construction.

**Figure 11 – Comparison between using Ensembl and OrthoMCL in constructing the interolog network**

Ensembl [17] contains protein orthologs based on sequence similarity as well as domain information, while OrthoMCL [18] is predominantly based on sequence similarity. As we can see from the table, using domain information (through Ensembl) generates significantly more many-to-many ortholog mappings thereby enhancing our interolog construction.

## Tables

**Table 1 – Properties of yeast physical PPI datasets**

| Database | # proteins | # (non self and duplicated) interactions |
|---|---|---|
| IntAct (version Nov 13, 2012) | 5276 | 18834 |
| Biogrid (version 3.2.95, Nov 30, 2012) | 5886 | 73923 |
| IntAct ∪ Biogrid | 6332 | 83777 |
| IntAct ∩ Biogrid | 4620 | 8930 |
| ICDScore(IntAct ∪ Biogrid) | 5239 | 71636 |

**Table 2 - Properties of human physical PPI datasets**

| Database | # proteins | #interactions |
|---|---|---|
| HPRD (Release 9, 2010) | 9617 | 39184 |
| Biogrid (April 25, 2012) | 12515 | 59027 |
| HPRD ∪ Biogrid | 13624 | 76719 |
| HPRD ∩ Biogrid | 8615 | 21491 |
| ICDScore(HPRD ∪ Biogrid) | 8521 (EntrezID) | 61868 |
| ICDEnrich(HPRD ∪ Biogrid) | 9764 (EntrezID) | 192053 (EntrezID) |

**Table 3 - Properties of manually curated protein complex datasets**

| Databases | # complexes |
|---|---|
| Wodak [28] yeast complexes (CYC 2008) | 149 with size>3 (36.5%) |
| | Total: 408 |
| CORUM [29] human complexes (September 2009) | 722 with size>3 (39.1%) |
| | Total: 1843 |

**Table 4 - Properties of the interolog network constructed from yeast and human PPIs**

| # Mapped nodes using orthology | 2470 |
|---|---|
| # Interologs | 6133 |
| Size of biggest connected component | 2434 nodes, 6112 edges |
| #Other connected components | 16 (size from 2-3) |

**Table 5 - Comparisons of different methods on yeast data**

Predicted complexes: resulting network clusters

Matched predictions: resulting network clusters that match with benchmarks

Precision = #matched prediction / #predicted complexes

Recall = # detected conserved complexes / # gold standard conserved complexes

| Method | # Predicted complexes | # Matched predictions | Precision | # Gold standard conserved complexes | # Detected conserved complexes | Recall (of conserved complexes) |
|---|---|---|---|---|---|---|
| COCIN | 71 | 36 | 50.7% | 42 | 32 | 76.2% |
| CMC | 1202 | 145 | 12.1% | 42 | 23 | 54.8% |
| HACO | 1040 | 69 | 6.6 % | 42 | 17 | 40.5% |
| MCL | 387 | 37 | 9.6% | 42 | 5 | 11.9% |

**Table 6 - Comparisons of different methods on human data**

Predicted complexes: resulting network clusters

Matched predictions: resulting network clusters that match with benchmarks

Precision = #matched prediction / #predicted complexes

Recall = # detected conserved complexes / # gold standard conserved complexes

One predicted complex of COCIN can match with many benchmark complexes, this explains for #detected conserved complexes > #matched predictions (as illustrated in **Figures 5-8**)

| Method | # Predicted complexes | # Matched predictions | Precision | # Gold standard conserved complexes | # Detected conserved complexes | Recall (of conserved complexes) |
|--------|----------------------|----------------------|-----------|-------------------------------------|-------------------------------|--------------------------------|
| COCIN  | 71                   | 36                   | 50.7%     | 118                                 | 78                            | 66.1%                          |
| CMC    | 1389                 | 156                  | 11.2%     | 118                                 | 66                            | 55.9%                          |
| HACO   | 1290                 | 80                   | 6.2%      | 118                                 | 36                            | 30.5%                          |
| MCL    | 631                  | 45                   | 7.1%      | 118                                 | 24                            | 20.3%                          |

**Table 7 – Additional conserved complexes found in yeast**

| ID | Complex name | Size | Jaccard score | Functional category | Functional description |
|---|---|---|---|---|---|
| 96 | eIF3 complex | 7 | 0.63 | Translation | Eukaryotic translation initiation factor |
| 247 | Transcription factor TFIID complex | 15 | 0.73 | Transcription | mRNA synthesis |
| 27 | DNA-directed RNA polymerase II complex | 12 | 0.69 | Transcription | mRNA synthesis |
| 45 | DNA replication factor C complex (Rad24p) | 5 | 0.67 | DNA processing | DNA synthesis and replication |
| 152 | DNA replication factor C complex (Rcf1p) | 5 | 0.67 | DNA processing | DNA synthesis and replication |
| 294 | Mcm2-7 complex | 6 | 0.6 | DNA processing | Chromosome maintainance, DNA synthesis and replication |
| 268 | SF3b complex | 6 | 0.57 | RNA processing | mRNA splicing |
| 65 | U6 snRNP complex | 8 | 0.5 | RNA processing | This complex combines with other snRNPs, unmodified pre-mRNA, and various other proteins to assemble a spliceosome, a large RNA-protein molecular complex upon which splicing of pre-mRNA occurs. |
| 375 | AP-3 adaptor complex | 4 | 0.67 | Cellular transport, vesicular transport | This complex is responsible for protein trafficking to lysosomes and other related organelles. |
| 25 | 20S proteasome | 14 | 0.5 | Cell cycle, protein fate | Proteasomal degradation (ubiquitin/proteasomal pathway), protein processing (proteolytic) |
| 137 | Chaperonin-containing T-complex | 8 | 0.67 | Protein fate | A multisubunit ring-shaped complex that mediates protein folding in the cytosol without a cofactor. |

## Table 8 – Additional conserved complexes found in human

| ID | Complex name | Size | Jaccard score | Functional category | Function description |
|---|---|---|---|---|---|
| 4392 | EIF3 complex (EIF3A, EIF3B, EIF3G, EIF3I, EIF3C) | 5 | 0.57 | Translation | Translation initiation |
| 4403 | EIF3 complex (EIF3A, EIF3B, EIF3G, EIF3I, EIF3J) | 5 | 0.57 | Translation | Translation initiation |
| 104 | RNA polymerase II core complex | 12 | 0.69 | Transcription | mRNA synthesis |
| 2685 | RNA polymerase II | 17 | 0.59 | Transcription | mRNA synthesis |
| 2686 | BRCA1-core RNA polymerase II complex | 13 | 0.64 | Transcription | mRNA synthesis |
| 471 | PCAF complex | 10 | 0.6 | Transcription, DNA processing | DNA conformation modification (e.g. chromatin), modification by acetylation, deacetylation, organization of chromosome structure. |
| 2200 | RFC2-5 subcomplex | 4 | 0.5 | DNA processing | DNA synthesis and replication |
| 387 | MCM complex | 6 | 0.6 | DNA processing | Chromosome maintainance, DNA synthesis and replication |
| 369 | MMR complex 2 | 4 | 0.67 | DNA processing | DNA damage repair |
| 290 | MSH2-MLH1-PMS2-PCNA DNA-repair initiation complex | 4 | 0.67 | DNA processing | DNA damage repair initiation |
| 1169 | SNARE complex | 4 | 0.6 | Cellular transport, vesicular transport | Vesicle fusion, synaptic vesicle exocytosis |
| 562 | LSm2-8 complex | 7 | 0.67 | RNA processing | mRNA splicing |
| 561 | LSm1-7 complex | 7 | 0.67 | RNA processing | Control of mRNA stability during splicing |
| 3036 | Ubiquitin E3 ligase (SKP1A, SKP2, CUL1, CKS1B, RBX1) | 5 | 0.5 | Cell cycle, protein fate | Mitotic cell cycle and cell cycle control, modification by ubiquitination, deubiquitination |
| 2188 | Ubiquitin E3 ligase (CDC34, NEDD8, BTRC, CUL1, SKP1A, RBX1) | 5 | 0.5 | Cell cycle, protein fate | Mitotic cell cycle and cell cycle control, modification by ubiquitination, deubiquitination |
| 2189 | Ubiquitin E3 ligase (SMAD3, BTRC, CUL1, SKP1A, RBX1) | 5 | 0.5 | Cell cycle, protein fate | Mitotic cell cycle and cell cycle control, modification by ubiquitination, deubiquitination |

**Table 9 – Details of gold standard testing dataset for conserved protein complexes between human and yeast**

| Score usage | MSJ ≥threshold |
|---|---|
| Threshold | 50% |
| # conserved yeast complexes | 42/149 with size>3 (28.1%) |
| | Total: 79/408 (19.3%) |
| # conserved human complexes | 118/722 with size>3 (16.3%) |
| | Total: 219/1843 (11.9%) |

**Table 10 - Homology data: Ensembl and OrthoMCL**

Ensembl [17] contains protein orthologs based on sequence similarity as well as domain information, while OrthoMCL [18] is predominantly based on sequence similarity. As we can see from the table, using domain information (through Ensembl) generates significantly more many-to-many ortholog mappings thereby enhancing our interolog construction.

| | | Ensembl database | OrthoMCL database |
|---|---|---|---|
| # Ortholog groups: | # 1-to-1 groups | 1096 | 1153 |
| | # 1-Yeast-to-many groups | 756 | 434 |
| | # 1-Human-to-many groups | 116 | 116 |
| | # many-to-many groups | 197 | 167 |
| | Total: | 2165 (5503 pairs) | 1870 |
| # Human paralog groups: | | 2573 | 2435 |
| # Yeast paralog groups: | | 426 | 393 |
| Total # homolog groups: | | 5164 | 4698 |

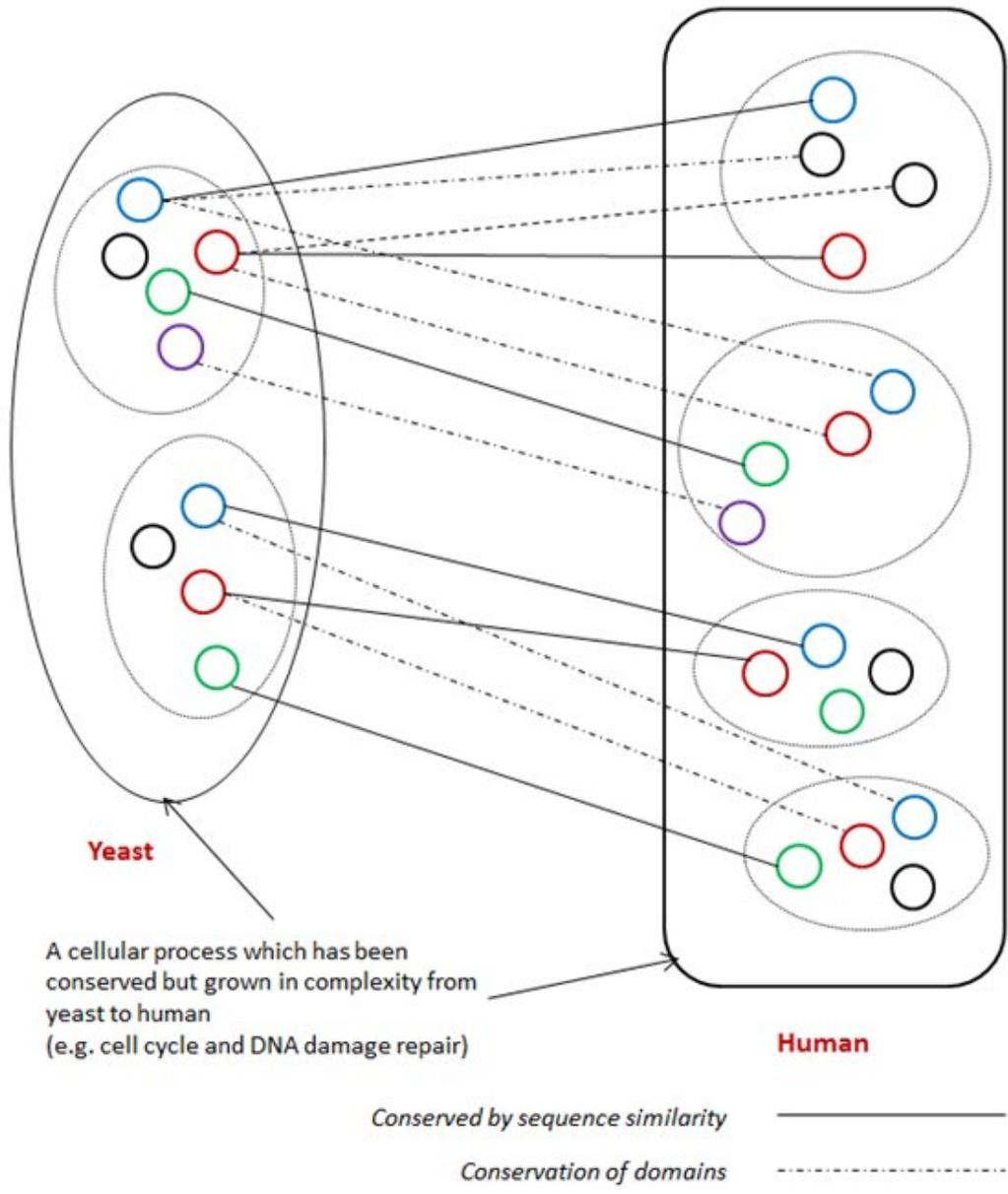

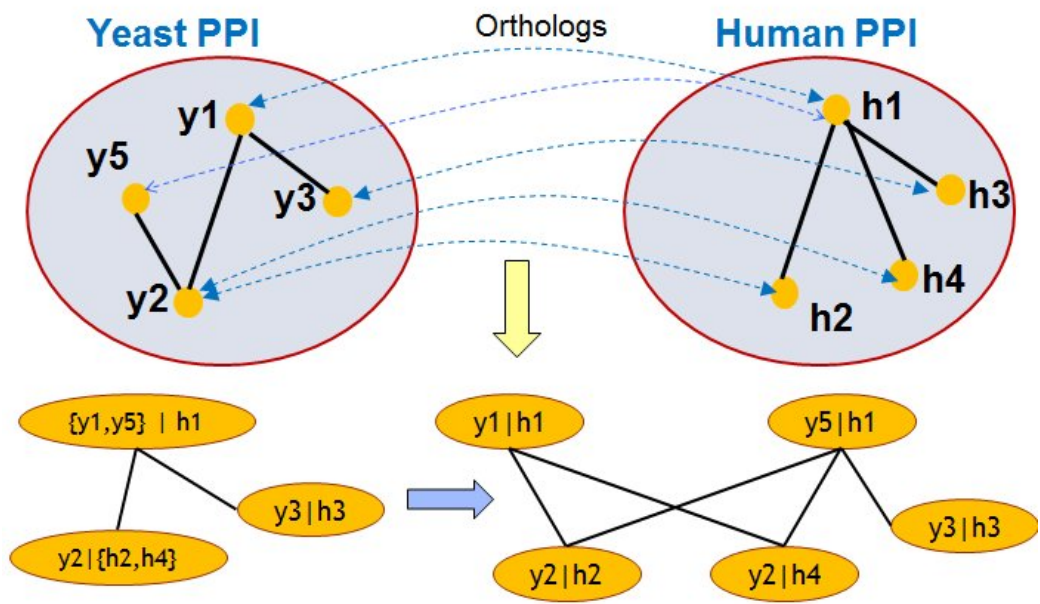

- **Conservation scores:**

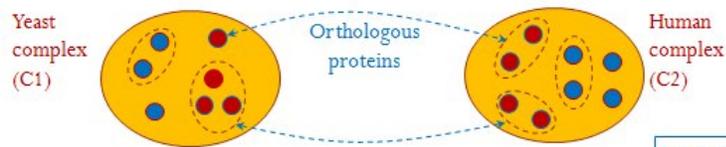

Yeast complex (C1) — Orthologous proteins — Human complex (C2)

- $I_{C_i}(g_i)$: #proteins of group $g_i$ in complex $C_i$.
- $G_{C_i}$: set of orthoGroups in complex $C_i$.

- Multi-set Jaccard score:

$$\mathrm{MSJ}(C_1, C_2) = \frac{\sum_{g_i \in (G_{C_1} \cup G_{C_2})} \min(I_{C_1}(g_i), I_{C_2}(g_i))}{\sum_{g_i \in (G_{C_1} \cup G_{C_2})} \max(I_{C_1}(g_i), I_{C_2}(g_i))} = \frac{\min(1,2) + \min(3,2)}{7 + \max(1,2) + \max(3,2)} = 0.25$$

- $0 \leq \mathrm{MSJ} \leq 1 \; (\forall C_1, C_2)$

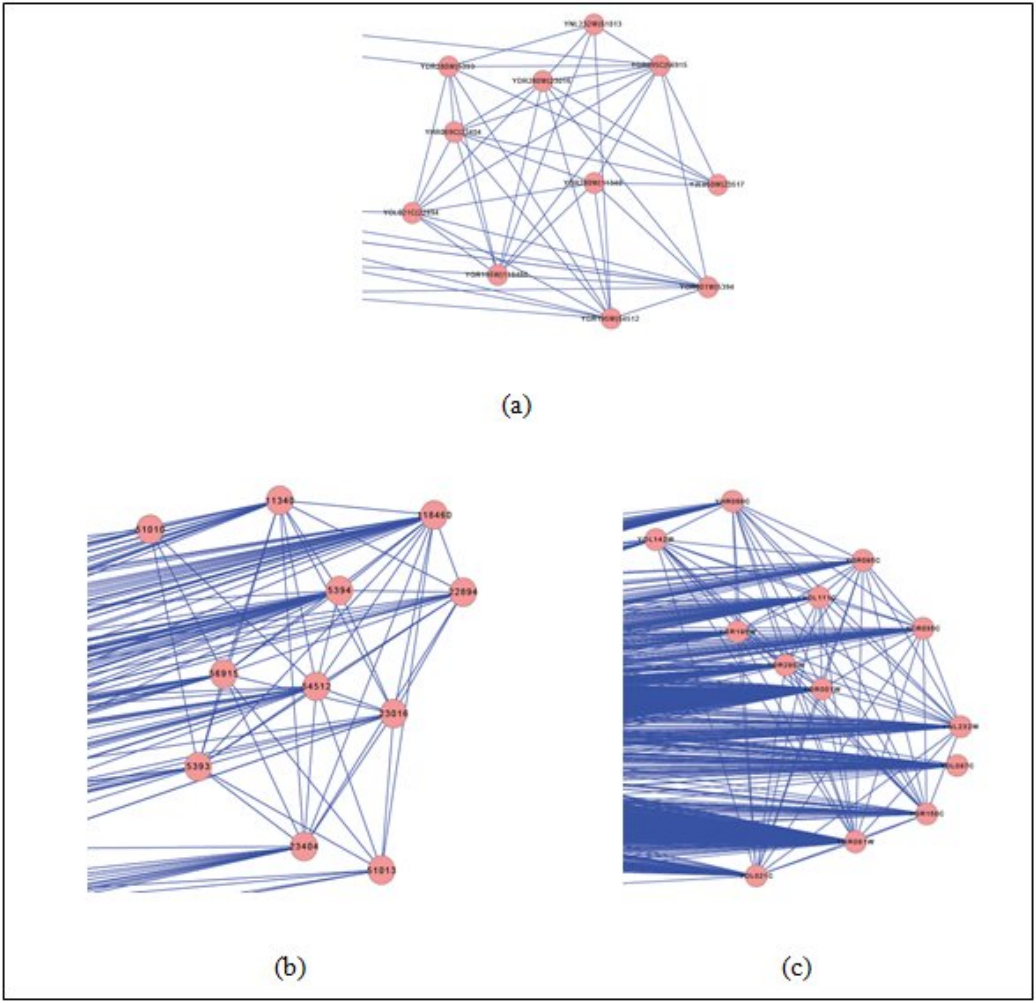

(a)

(b) (c)

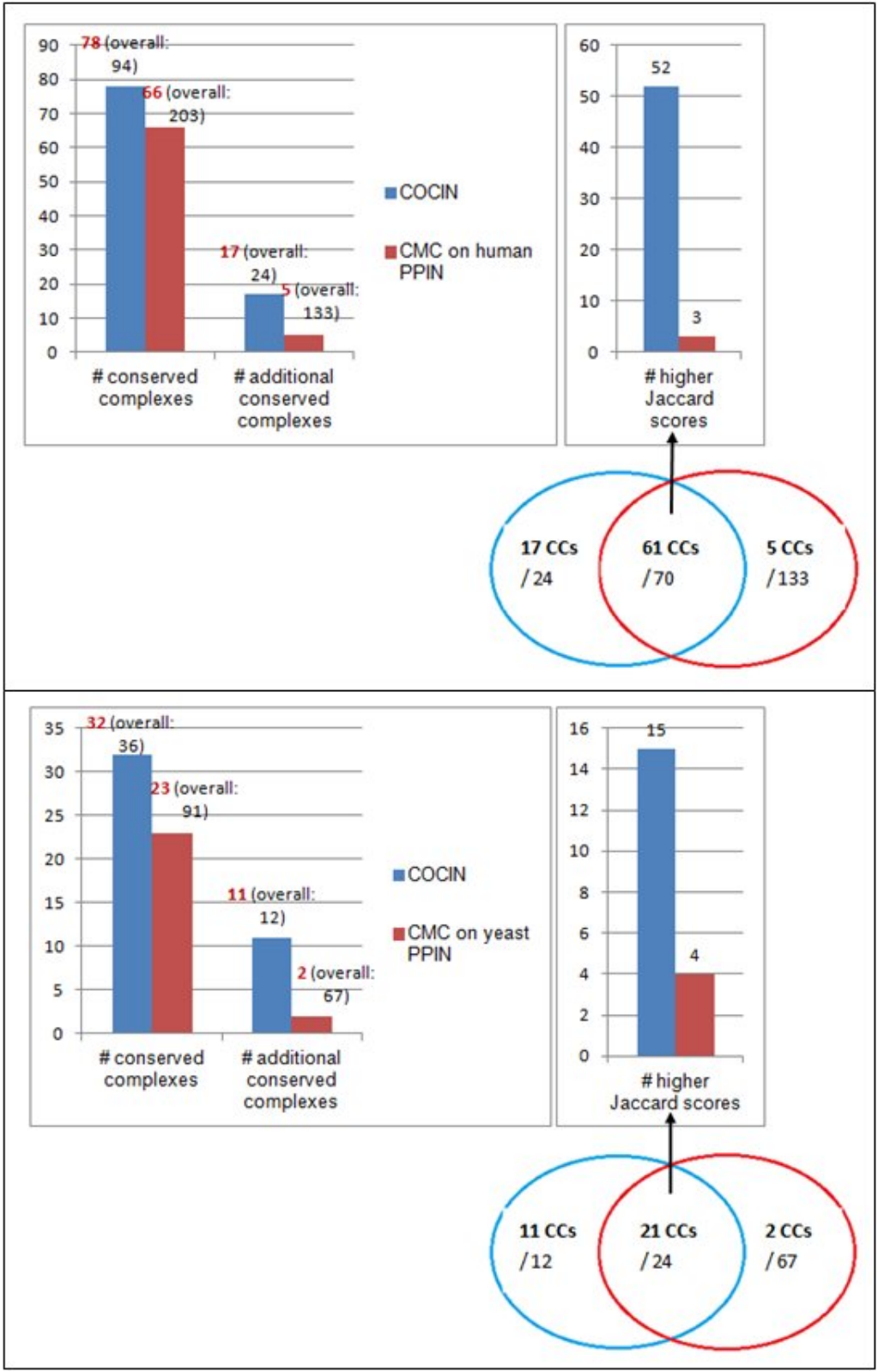

- **Example 1:** Human RNA Polymerase II core complex (complex ID= 104): {5430, 5431, ..., 5441}

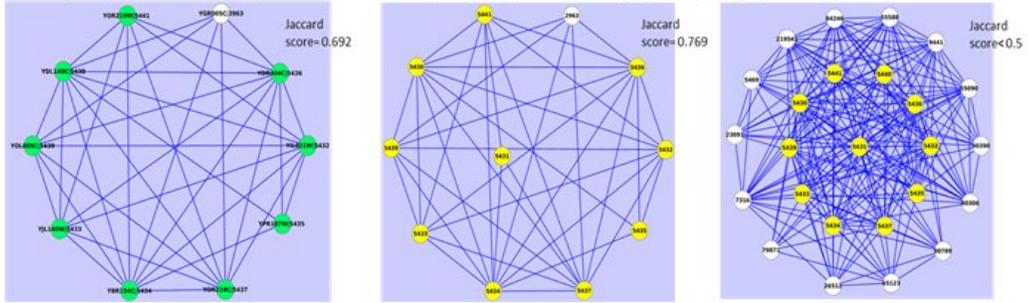

- **Example 2:** Human LSm1-7 complex (complex ID = 561): {27257, 57819, 27258, 25804, 23658, 11157, 51690, 51691}

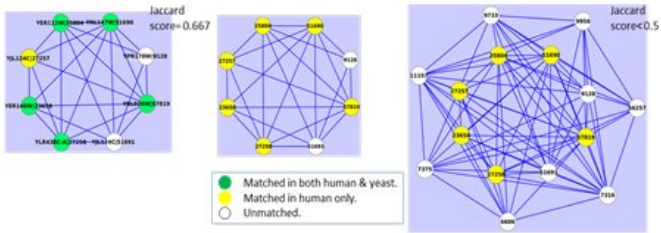

- 🟢 Matched in both human & yeast.
- 🟡 Matched in human only.
- ⚪ Unmatched.

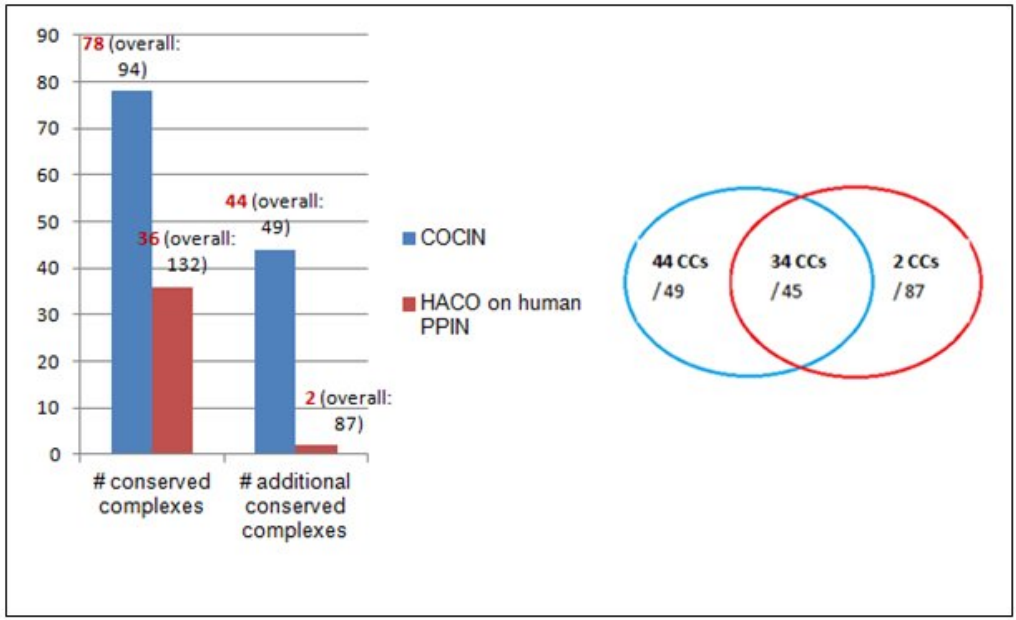

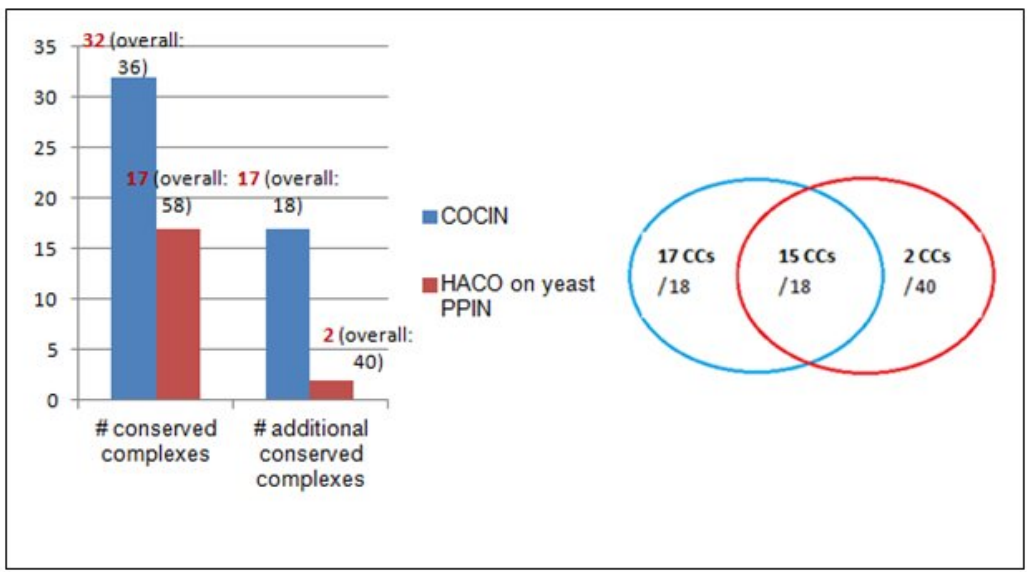

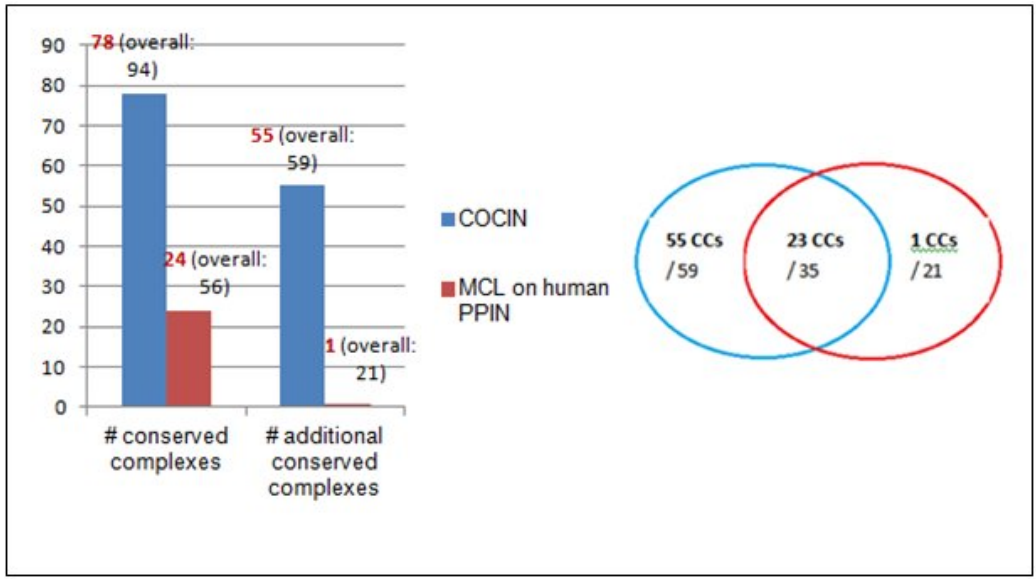
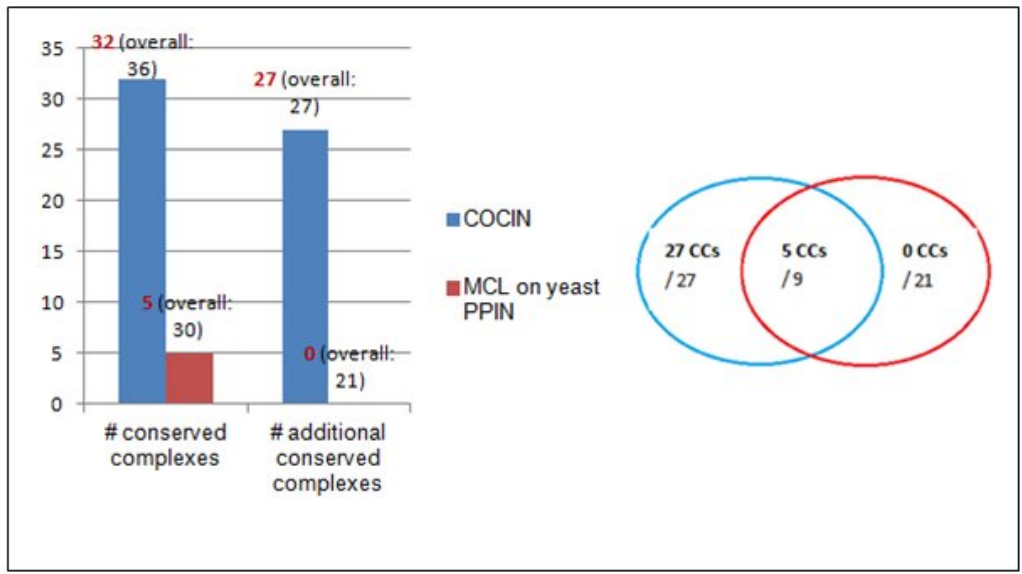

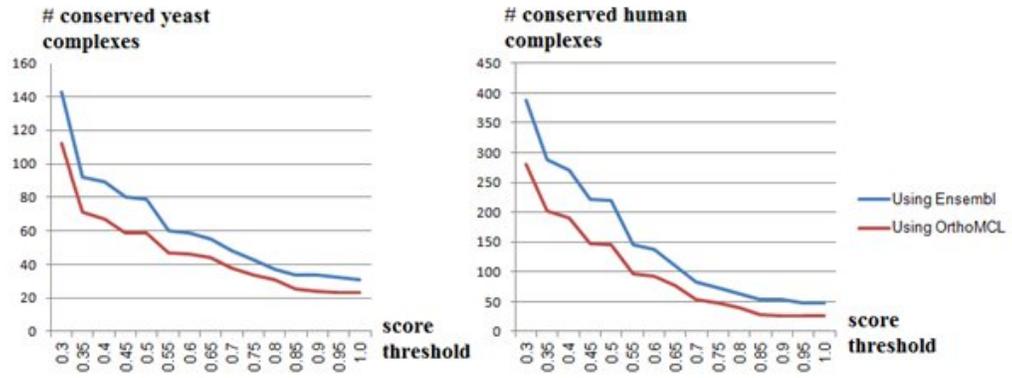

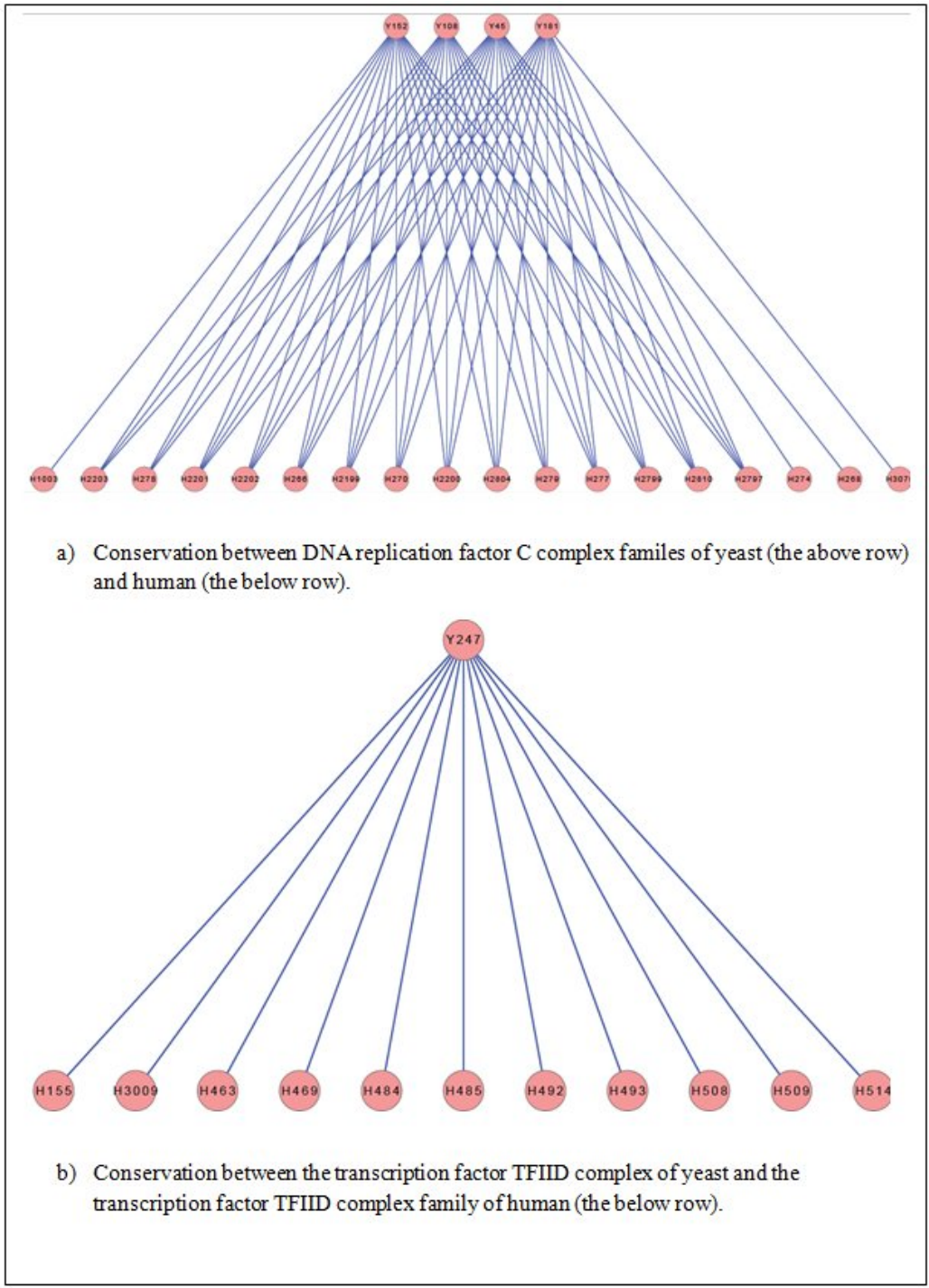

a) Conservation between DNA replication factor C complex familes of yeast (the above row) and human (the below row).

b) Conservation between the transcription factor TFIID complex of yeast and the transcription factor TFIID complex family of human (the below row).

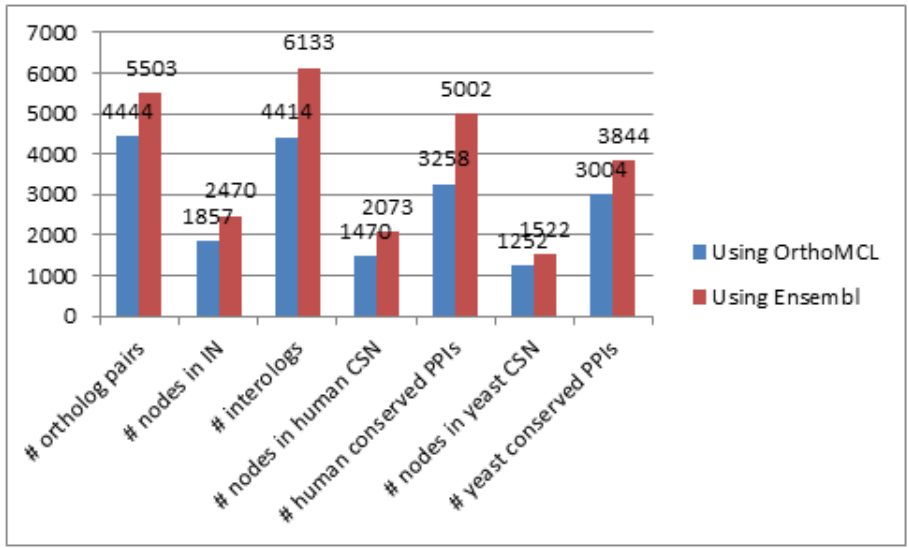